# Mass Spectrometry Studies of Hydrogen Ions Energy Distributions in an ECR- based Large Volume Plasma Source

Bibekananda Naik[1], Ramesh Narayanan[1], Debaprasad Sahu[1],

Mainak Bandyopadhyay[2,3], Ashish Ganguli[1]

[1] Indian Institute of Technology Delhi, Hauz Khas, New Delhi, Delhi 110016, India

[2] Institute for Plasma Research, Gandhinagar, Gujarat 382428, India

[3] Homi Bhabha National Institute, Mumbai, Maharashtra 400094, India

Email ID: bibekanandanaik74@gmail.com, rams@dese.iitd.ac.in, dpsahu@dese.iitd.ac.in

**Abstract:** Plasma is produced in a Large Volume Plasma Source (LVPS; dia. ≈ 1 m, height ≈ 1m) using CW microwaves (≈ 400 – 600 W, 2.45 GHz), in a compact ECR plasma source (CEPS) attached to LVPS, at hydrogen gas pressures ≈ 1 – 3 mTorr. Plasma expands along the CEPS magnetic field into LVPS. A Hiden Analytical HPR 60 molecular beam mass spectrometer (MBMS) is used to measure the $H^-$ ion energy distribution functions (IEDFs) in the downstream plasma. Previous plasma characterization studies in LVPS indicated favourable downstream plasma conditions for volume production of $H^-$ ions. Measurements conducted with the MBMS probe aligned facing the plasma flow ≈ 80 cm downstream, gave typical $H^-$ count rates $\approx 3 \times 10^5$ counts /s, at ≈ 400 W, ≈ 1 mTorr, along with a distinct high energy tail ($\lesssim 20$ eV). These and other results are analyzed in detail. The positive ion spectrum showed the $H_3^+$ count to be consistently high in all cases ($\approx 60 - 70$ %); the counts for $H_2^+$ and $H^+$ were $\approx 30 - 35$ % and a ≈ few %. Combining the Langmuir probe (LP) and MBMS data it is possible to determine the approximate densities in front of the MBMS probe aperture. At ≈ 500 W and ≈ 2 mTorr, one finds: $n_{H^+} \approx 9.6 \times 10^9$ $\text{cm}^{-3}$, $n_{H_2^+} \approx 1.7 \times 10^{10}$ $\text{cm}^{-3}$ and $n_{H_3^+} \approx 4.3 \times 10^{10}$ $\text{cm}^{-3}$. The corresponding $H^-$ density, ≈ 80 cm downstream is $n_{H^-} \approx 3.9 \times 10^8$ $\text{cm}^{-3}$. Accounting for all $H^-$ losses due to scattering and destruction, one finds the effective mean free path for $H^-$ loss to be ≈ 12.4 cm. Noting that $H^-$ formation takes place about ≈ 10 - 30 cm downstream of the source exit, the approximate average $H^-$ density in the formation zone is determined as $\approx 5.5 \times 10^{10}$ $\text{cm}^{-3}$. This value is remarkably encouraging for $H^-$ production in volume mode, considering the large chamber volume and area, as well as the very moderate power used for the experiments.

## 1. Introduction

Hydrogen plasma discharges are among the most extensively investigated plasma systems owing to their wide range of applications in diverse technological domains. Hydrogen being a molecular gas, the coexistence of multiple ionic species such as $H^+$, $H_2^+$, $H_3^+$, and negative hydrogen ions ($H^-$) results in complex plasma chemistry and kinetics that are unique to hydrogen discharges. A distinctive feature of hydrogen plasmas is the strong influence of vibrationally excited $H_2$ molecules and electronically excited H atoms on ionization balance, dissociation pathways, and overall plasma chemistry. Hydrogen plasmas are important for many applications covering the field of material processing [1], production of ion beams [2], accelerator physics [3] and thermonuclear fusion [4].

Negative ions ($H^-$) generated in the hydrogen plasmas under certain working conditions find their applications in various fields ranging from microelectronics [5] to magnetically confined thermonuclear fusion [4], as sources for particle accelerators [6] to space propulsion [7]. In thermonuclear fusion, $H^-$ ions are vital for the Neutral Beam Injection (NBI) heating and current drive of fusion plasmas thereby facilitating significant fusion reactions. The negative ions are preferred over positive ions owing to their higher neutralization efficiency at the required high beam energy of hundreds of keV to MeV [8]. The International Thermonuclear Experimental Reactor (ITER) is designed to employ two negative ion–based neutral beam injection (NBI) systems, each capable of delivering 17 MW of heating power for pulse durations of up to one hour [9]. These injectors require $H^-$ ion beams with total currents of approximately 45 A, accelerated to energies of about 1 MeV and subsequently neutralized in a gas cell before injection into the tokamak plasma. Achieving the required beam performance necessitates the extraction of $H^-$ ions at current densities of approximately 30–35 mA $cm^{-2}$. This, in turn, demands the generation of a highly uniform plasma (uniformity ~10%) with an electron density of $\sim 10^{12}$ $cm^{-3}$ and an electron temperature of ~1 eV in front of the extraction grid over a large area of approximately 2.0 m × 0.6 m [10]. In addition to the heating, ITER will employ a diagnostic neutral beam (DNB) system based on $H^-$ ions, designed to produce 100 keV neutral beams with currents of about 60 A for plasma diagnostics [11].

To meet the requirements of ITER and future fusion reactors, significant research efforts have been devoted to the development of high-current negative ion sources. At present, the most widely adopted technology is based on radio-frequency (RF) driven inductively coupled plasmas (ICPs), operating at power levels approaching 1 MW [10]. In these sources, energy transfer from the RF field to the plasma occurs primarily through collisional processes, which are generally less efficient than resonant heating mechanisms. Electron cyclotron resonance (ECR) heating, on the other hand, offers highly efficient resonant power coupling and has the potential to generate plasmas with superior efficiency and uniformity. Despite these advantages, the application of ECR technology to large-area, high-current negative ion sources remains relatively unexplored owing to the scientific and engineering challenges associated with scaling ECR plasmas to reactor-relevant dimensions.

In low-pressure plasmas, $H^-$ ions are mainly produced through two mechanisms: surface production and volume production. In the surface mode, the plasma-facing surface is coated with a low-work-function material, such as Cesium, which greatly enhances $H^-$ emission when hydrogen atoms/ions strike the surface. However, Cs-based $H^-$ source operation has major demerits such as high chemical reactivity, surface contamination/instability, frequent reconditioning, and serious maintenance and safety issues, leading to poor long-term reliability [12]. In contrast, volume production relies on plasma reactions (mainly involving vibrationally excited $H_2$), avoiding dependence on unstable surface conditions. Therefore, exploring and optimizing volume-mode $H^-$ production is essential for achieving Cs-free, stable, and efficient long-duration negative ion sources.

In the source reported here, the negative hydrogen ions are produced via volume mode which involves dissociative electron attachment of slow electrons (1 - 2 eV) to the vibrationally excited hydrogen molecules, the latter being produced primarily by fast electrons (15 - 40 eV) colliding with ground state hydrogen molecules [13]. The efficiency of this process is strongly governed by the population of vibrationally excited molecules and the availability of low-energy electrons in the plasma. At the same time, the survival of $H^-$ ions in the plasma volume is limited by various destruction processes, such as electron detachment and other collisional losses, which significantly influence the overall negative ion density. Therefore, understanding both the production and loss mechanisms is essential for controlling $H^-$ yield in the source. In this context, ion energy distribution functions (IEDFs) of $H^-$ ions act as a critical diagnostic.

The present study focuses on molecular beam mass spectrometric (MBMS) measurements of IEDFs of $H^-$ ions in an ECR hydrogen plasma source developed for large-area $H^-$ ion production for fusion applications. Plasma produced in a Compact ECR Plasma Source [14] using CW microwaves at 2.45 GHz at ≈ 400 – 600 W of power at ≈ 1- 3 mTorr pressure, is allowed to flow along the CEPS magnetic field into the large volume expansion chamber attached to the CEPS. IEDF measurements were conducted in two configurations, with the MBMS probe assembly aligned transverse-to and in-line with the plasma flow. The results are analyzed in detail. In the in-line configuration the plasma sampling probe of the MBMS is ≈ 80 cm downstream from the CEPS source exit *facing* the flowing plasma. Using MBPS counts for positive ions and combining the results with Langmuir probe (LP) and the IEDF measurements it is possible to determine the densities of the $H_3^+$, $H_2^+$, $H^+$ ions in front of the MBMS probe and subsequently, one may calculate the $H^-$ ion density as well. Regarding the scattering and destruction of the $H^-$ ions as random collisional, it is possible to determine the effective mean free path for $H^-$ ion loss as these travel towards the MBMS probe. Noting additionally that the $H^-$ ion production site is about ≈ 10 – 30 cm downstream from the CEPS source, it is possible to determine the average density of the $H^-$ ions in their formation zone. This turns out to be fairly promising the size of the chamber and the modest microwave power used. To the best of the authors' knowledge, such measurements identifying the nature of $H^-$ energy distributions in a large-area negative ion beam source have not been reported elsewhere.

The paper has five sections. Section 2 describes the experimental system, including the compact ECR plasma source (CEPS) and the Molecular Beam Mass Spectrometer (MBMS) [Model: Hiden HPR-60] utilized for the IEDF measurement. The IEDF results of negative hydrogen ions under various operating experimental conditions for two source configurations are presented in section 3. Section 4 presents an analysis and discussion of the results. In particular, the densities of the different species (positive and negative) are determined, along with the density of the $H^-$ ions in the formation zone. Section 5 summarizes the results of the paper. The paper includes two appendices. *Appendix A* provides the detailed calculations for the parallel energy gain of $H^-$ ions during transport along the diverging magnetic field lines and *Appendix B* gives the methodology for combining the LP and MBMS results for determining the different ion densities.

## 2. Experimental Setup and Plasma Diagnostic

### 2.1 Expansion Chamber and Microwave Plasma Source

Experiments were conducted in an ECR based large area negative ion beam source (ELNIBS) a schematic of which is shown in Fig. 1. The setup consists of a stainless-steel (ss), cylindrical expansion chamber (≈ 1 metre in height and diameter) mounted vertically on the support structure (not shown in the figure). The expansion chamber is assembled by stacking multiple cylindrical sections [i.d. ≈ 1 m, height ≈ 6 inches] vertically one above the other, between two hemispherical domes at the top and bottom [15]. The top dome is fitted with several ports: one source port at the centre on top of the dome and the others placed around the central port. In the experiments presented in this paper, the CEPS was mounted on the central port, while the remaining ports provided access for the various plasma diagnostics. While the ports on the top dome provide vertical access to the plasma and the diagnostics, the vertically stacked cylindrical sections also have ports that for mounting sources and sensors / diagnostics that provide radial access to the plasma and the diagnostics. The bottom dome has a central flange through which the chamber is pumped. A base pressure of $\approx 5 \times 10^{-6}$ Torr is attained, which is ≈ 3 orders below the working pressure. The experiments were conducted with hydrogen as the feed gas, with the working pressure varied in the range, $\approx 1 - 5$ mTorr. More details of the expansion chamber are presented in Ref. [16]. Mass spectrometry studies of the plasma were conducted with the help of Hiden HPR-60 MBMS. Its plasma sampling mass spectrometer (PSMS) probe was mounted on the expansion chamber through one of the ports available on a cylindrical section. More information about the mounting and the operation of the mass spectrometer is presented later.

The indigenous, Compact ECR Plasma Source (CEPS), a patented source [17] of Plasma Lab at IIT Delhi (PPL – IITD) weighs only 14 kg (including the NdFeB ring magnets) and has an overall length of ≈ 60 cm. The CEPS comprises a magnetron [≈ 800 W, 2.45 GHz, CW] for launching microwaves into a rectangular waveguide [WR340; TE10 mode] followed by a dual directional coupler (for monitoring power), a triple stub impedance matching tuner and a rectangular to cylindrical guide transducer. The latter feeds the microwave power into the plasma source section (PSS), which is a water cooled, cylindrical ss section (i.d. ≈ 9.1 cm, length ≈ 11 cm) on which the ring magnets are mounted coaxially. Vacuum integrity in the PSS is maintained by a quartz window seal between the rectangular-to-circular guide transducer and the PSS. The NdFeB ring magnets produce an ECR zone within the PSS where the microwave power is absorbed [Fig 1]. The ECR surface ($B \approx 875$ G, corresponding to an ECR frequency ≈ 2.45 GHz) within the PSS has a very complex magnetic field topology, with a magnetic null on the axis close to the source exit. A detailed account of the working of the CEPS and the action of the magnets can be found in Ref. [14, 18].

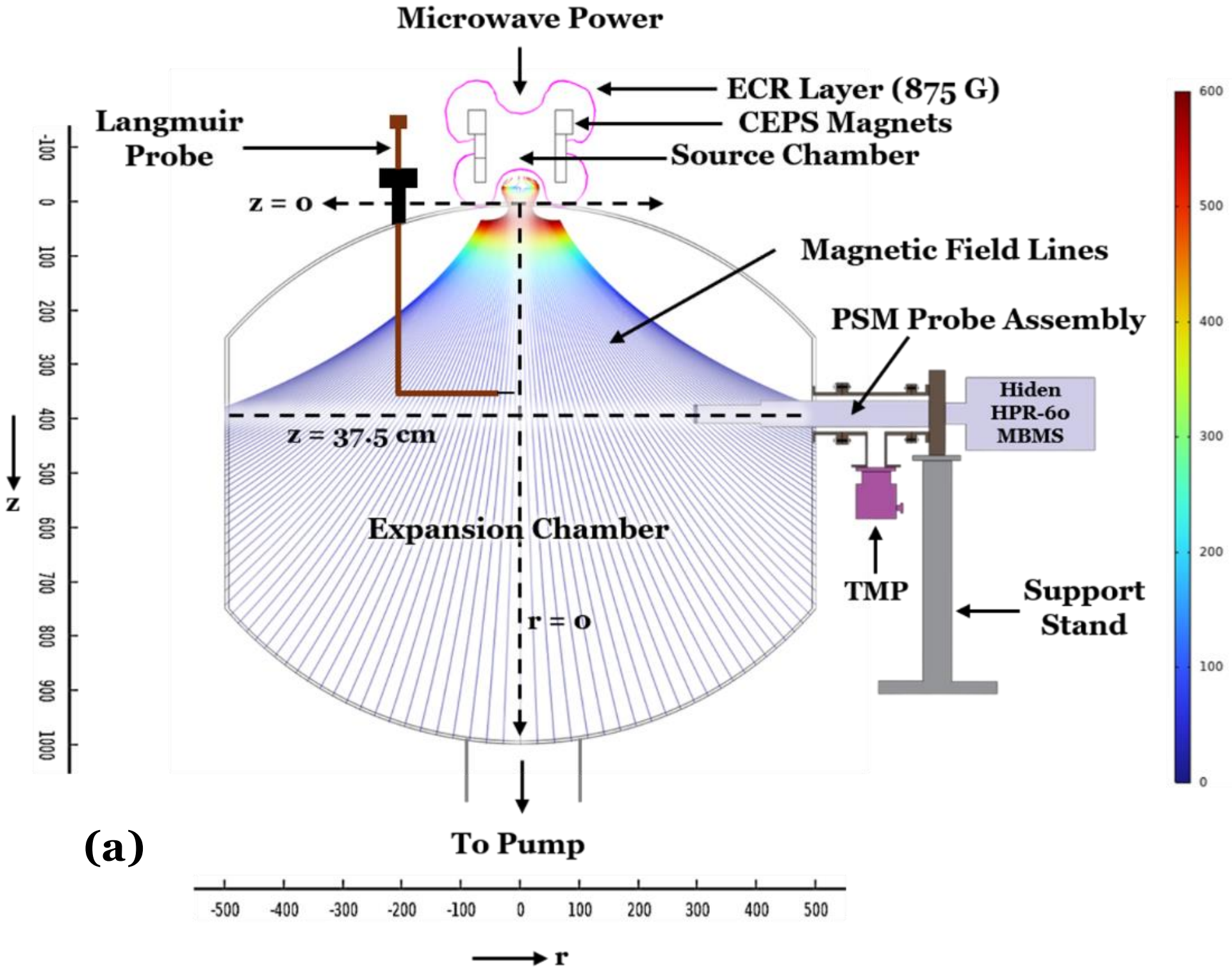

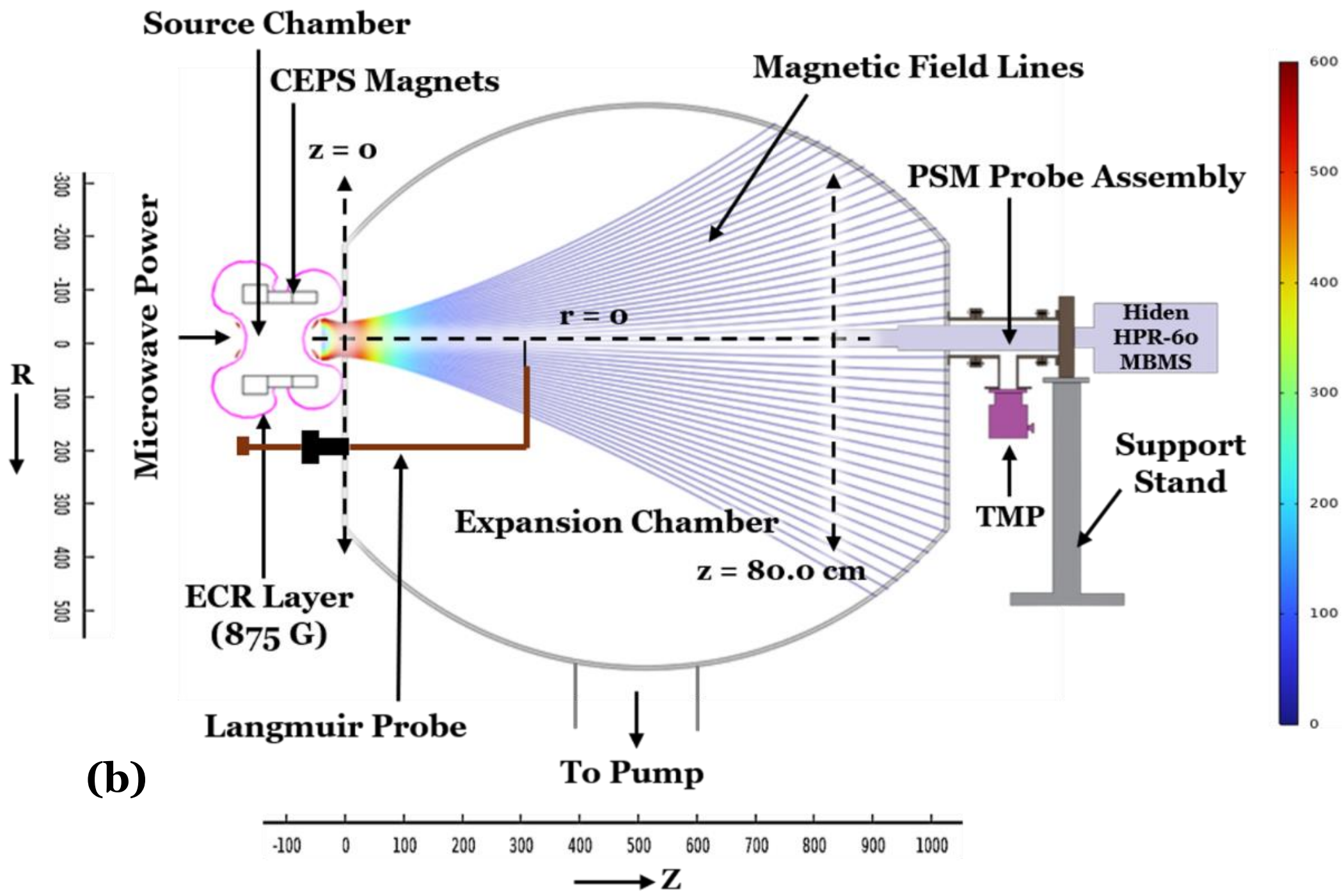

Figure 1. Experimental setup: Configuration A: CEPS at the top dome of the expansion chamber (a) and Configuration B: CEPS facing MBMS (b) (Size not to scale)

The magnets produce a diverging magnetic field within the larger expansion chamber, facilitating plasma flow. Impedance matching using the triple stub tuner, was manually controlled to keep the reflected microwave power within $\approx 10$ % of the forward power, during the operation of the source. The studies reported in this work were conducted using only a single CEPS at a time. However, as can be seen from the figure 1, experiments were conducted in two different source configurations. ***Configuration A:*** The CEPS was mounted on the central port of the top dome, allowing the plasma to flow vertically downwards into the expansion chamber, *transverse to the axis of the PSMS*, but *in-line* with the pumping port (Fig. 1(a)); ***Configuration B:*** The CEPS was mounted onto a source port on the *sidewall of a cylindrical section*, allowing the plasma to flow in *radially*, *in-line with the PSMS axis*, but *transverse* to the pumping port (Fig. 1(b)). Thus, the two configurations generate different *relative orientations* between the direction of plasma flow (and magnetic field axis) and the direction of pumping that sets up an additional drift. As a result, characterization of the *molecular plasma* in the two configurations is expected to throw light on the different charge and molecular ion states in the two orientations.

## 2.2 Molecular Beam Mass Spectrometer (MBMS)

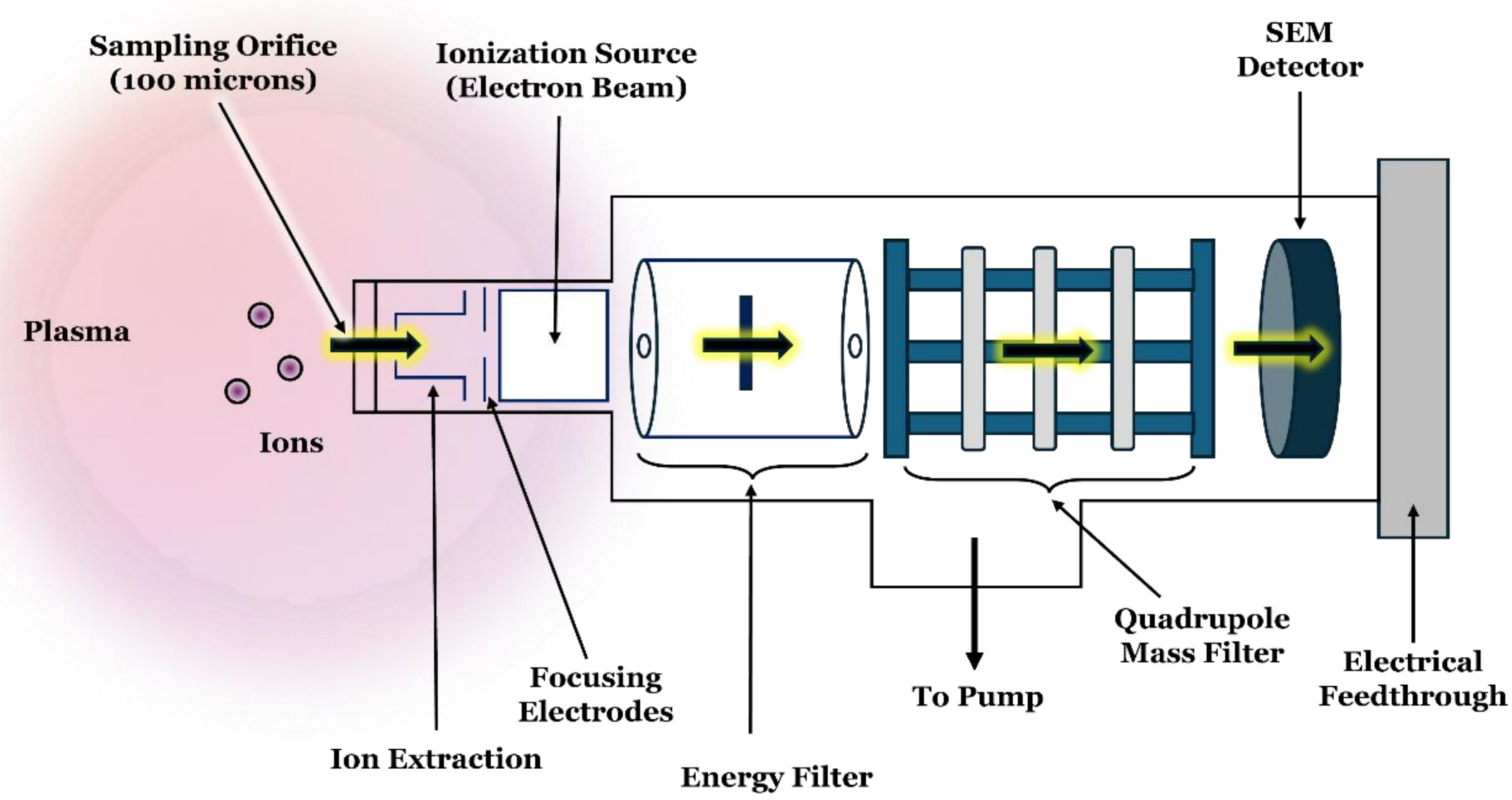

Figure 2. Schematic of the Hiden HPR-60 Molecular Beam Mass Spectrometer

As seen from Figs. 1, the orientation of the PSMS probe (relative to the chamber) remains the same in both configurations. It is seen from Fig. 1(a) that the arm of the PSMS is introduced radially in the $z \approx 37.5$ cm plane ($z = 0$ being at the exit plane of the CEPS). As already noted above, in *Configuration A*, the MBMS probe is transverse to the plasma flow *and* the magnetic field (on the $r = 0$ axis), while in *Configuration B*, the source and MBMS probe are *in-line* with both. Owing to mechanical constraints, the probe could only be inserted $\approx$ 20 cm inside the expansion chamber. Accordingly, in configuration A, MBMS measurements were performed at $r \approx 30$ cm, $z \approx 37.5$ cm and in configuration B, at $R \approx 0$, $Z \approx 80$ cm (the latter being read off on the attached ($R$, $Z$) scale). The schematic of the mass spectrometer probe assembly in shown in Fig. 2. The PSMS probe assembly is housed inside an ss tube of diameter, $\approx$ 4 cm. Plasma is sampled through the sampling orifice (size $\approx$ 100 micron) located at the centre of the endcap. The coordinates of the probe position as mentioned earlier correspond to the position of the orifice.

The mass spectrometer can be operated in the Residual Gas Analyzer (RGA), Positive (POS) or Negative (NEG) ion mode for detection of neutral atoms / molecules and ions, respectively. The PSMS probe assembly and the mass spectrometer unit consist of five sections: ion extraction and focusing, ionization source, energy filter, mass filter and the detector.

- ***Ion Extraction and Focusing***: The extraction optics consists of an extractor and an electrostatic lens used to extract and focus plasma ions onto the energy filter. These optics are placed before the ionization filament and can be biased to reject externally generated ions during neutral analysis.
- ***Ionization Source***: The ionization filament acts as an electron impact ionization source and is enabled only during the standard Residual Gas Analyzer (RGA) mode of operation. The generated ions are subsequently guided toward the energy filter using focusing electrodes.
- ***Energy Filter***: A Bessel Box energy filter is employed to selectively transmit ions of specific energies and is positioned between the ionization source and the mass filter. The system provides an energy resolution of ≈ 0.05 eV.
- ***Mass Filter***: The mass analysis is carried out using a triple-section quadrupole mass filter consisting of 9 mm molybdenum rods, providing a mass resolution of ≈ 0.02 amu.
- ***Detector***: Ion detection is performed using a Secondary Electron Multiplier (SEM) detector, along with an additional Faraday detector for measurements involving high count rates.

The PSMS probe assembly is differentially pumped via a turbo molecular pump, backed by a scroll pump, which ensures the safe operation of the probe components and the ion-counting detector. The pressure inside the probe housing is kept sufficiently low relative to the chamber pressure, enabling collision-free ion transport and proper operation of the SEM detector. During measurements, the vacuum base pressure inside the detector was maintained at ~ $10^{-6} - 10^{-8}$ Torr, below the hydrogen gas filling pressure ($\approx 10^{-3}$ Torr). The probe components were tuned following a standardized tuning procedure for both positive (POS) and negative (NEG) ion detection modes prior to the measurements. This involved optimizing the extraction optics, adjusting the bias voltages, and calibrating the mass spectrometer settings to ensure efficient transmission and accurate energy discrimination of the selected ion species. Particular care was taken to minimize signal distortion, suppress secondary electron effects, and maintain stable operating conditions, thereby ensuring reliable and reproducible IEDF measurements in both detection modes. Care was taken to minimize measurement artifacts and maintain stable operating conditions in both detection modes and the measurements were repeated over several detection cycles for error free and accurate results.

## 2.3 Source and Chamber Magnetic Field Configuration

As mentioned earlier, the magnetic field in the plasma source in the CEPS was generated using NdFeB ring magnets and is quite complex within the source chamber. Fig. 3(a) shows the simulated source magnetic field configuration generated in COMSOL Multiphysics [19]. The 2D contour plot of magnetic field intensity presented in the figure shows a single ECR layer (white contour) present inside the source section below the quartz window (in vacuum), that corresponds to a magnetic field strength of 875 G. (The remainder of the ECR contour is at atmospheric pressure and cannot initiate any plasma.) The on-axis profile of the magnetic field is shown in Fig. 3(b). The location of the ECR field and the on-axis null (at $z$ = - 23.0 mm) are also shown, with $z = 0$ corresponding to the PSS exit plane. Microwave power injected through the quartz window interacts with the plasma electrons that are confined by the mirror field (not shown here,) formed by the NdFeB ring magnets that allows electron transport from the axis to the wall and vice versa. In the process, the electrons pass through the resonant ECR surface, picking up energy from the microwave field. The on-axis null generates a separatrix that keeps the field before and after the null separate and aids the flow of electrons through the separatrix into the diverging field lines outside the PSS in the expansion chamber. A detailed discussion of the CEPS magnetic field and its action for plasma production is given in Ref. [18].

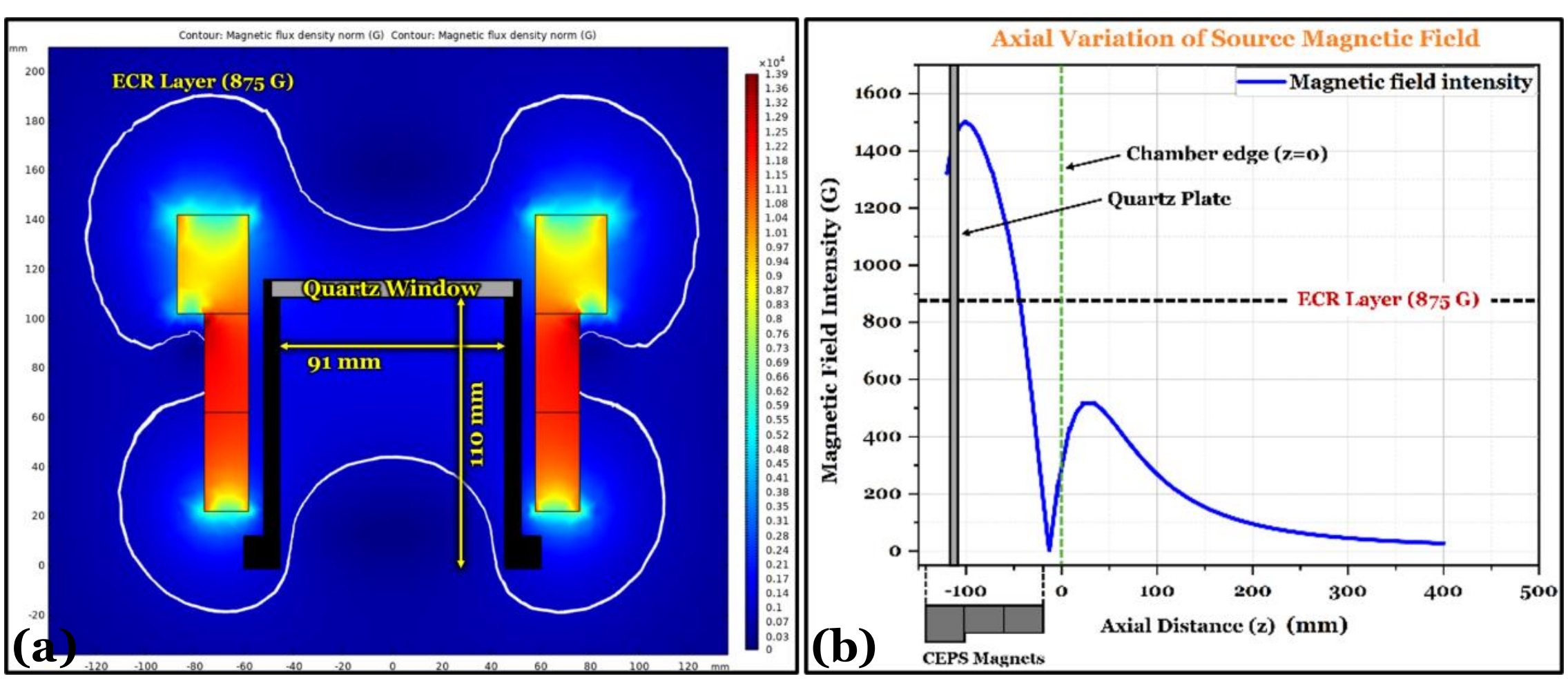

Figure 3. Contour plot of Magnetic field intensity showing a single ECR layers inside the source section (a); Axial variation of source magnetic field configuration (b)

The axial magnetic field profile shows an exponentially decaying field in the expansion chamber. The diverging field lines in the expansion chamber help produce uniform plasma at the downstream region of the expansion chamber. From the perspective of the mass spectrometer, the endcap is situated at a field strength of ≈ 13 G and ≈ 4 G for configuration A and B, respectively. These values are well below the allowed value of spectrometer operation i.e. ≈ 50 G, which ensures that the ambient magnetic field does not affect the MBMS measurements. Detailed experimental results in terms of negative ions detection and energy distributions are presented in the next section.

## 3. Experimental Observations

The volume mode production of $H^-$ ions is a two-step mechanism which involves (i) Vibrational excitation of hydrogen molecules by hot electrons (≈15 – 40 eV), and (ii) Dissociative attachment of slow electrons (≈ 0.5 – 4 eV) to the vibrationally excited hydrogen molecules for producing $H^-$ ions. From past studies [20, 21] conducted with the plasma source, it has been found that the plasma source produces a uniform hydrogen plasma in the downstream region of the expansion chamber ($z$ = 30 – 50 cm) over a cross-sectional area of diameter ~ 70 cm with density, $n_e \sim 10^{11}$ cm$^{-3}$ with low electron temperature $T_e \approx 1-2$ eV. Additionally, the presence of a hot electron population ($T_w \sim 50\ eV$) with lower density ($n_w \sim 10^8\ cm^{-3}$) have also been detected. Fig. 4 shows the radial and axial variation of the plasma parameters measured for configuration A and B, respectively that illustrates the behaviour of the plasma parameters as described. Measurements have also shown the presence of bulk electrons of comparable density with higher temperature $T_e >$ 20 eV close to the source mouth ($z$ = 0 cm), that lead to the formation of vibrationally excited Hydrogen molecules $H_2(v)$, the precursor of $H^-$ production in volume mode which are then transported from the upstream to the downstream region of the expansion chamber [22]. The plasma parameters achieved in the expansion chamber are favourable for volume production of $H^-$ ions. These encouraging conditions motivated the authors to investigate volume-produced $H^-$ ions using Molecular Beam Mass Spectrometry (MBMS), as reported in this paper, and to further explore the potential of the source as a negative ion beam source for large-area fusion applications. In this regard, experiments were conducted with Hydrogen at the operating pressure range of ≈ 1 – 3 mTorr and microwave powers ≈ 400 – 600 W. The mass spectrometer measurements were conducted to record the IEDF of $H^-$ ions for both source configurations, A and B. These are presented in Secs. 3.1 and 3.2 for Configurations A and Configuration B, respectively. A detailed comparison of the two configurations is provided later.

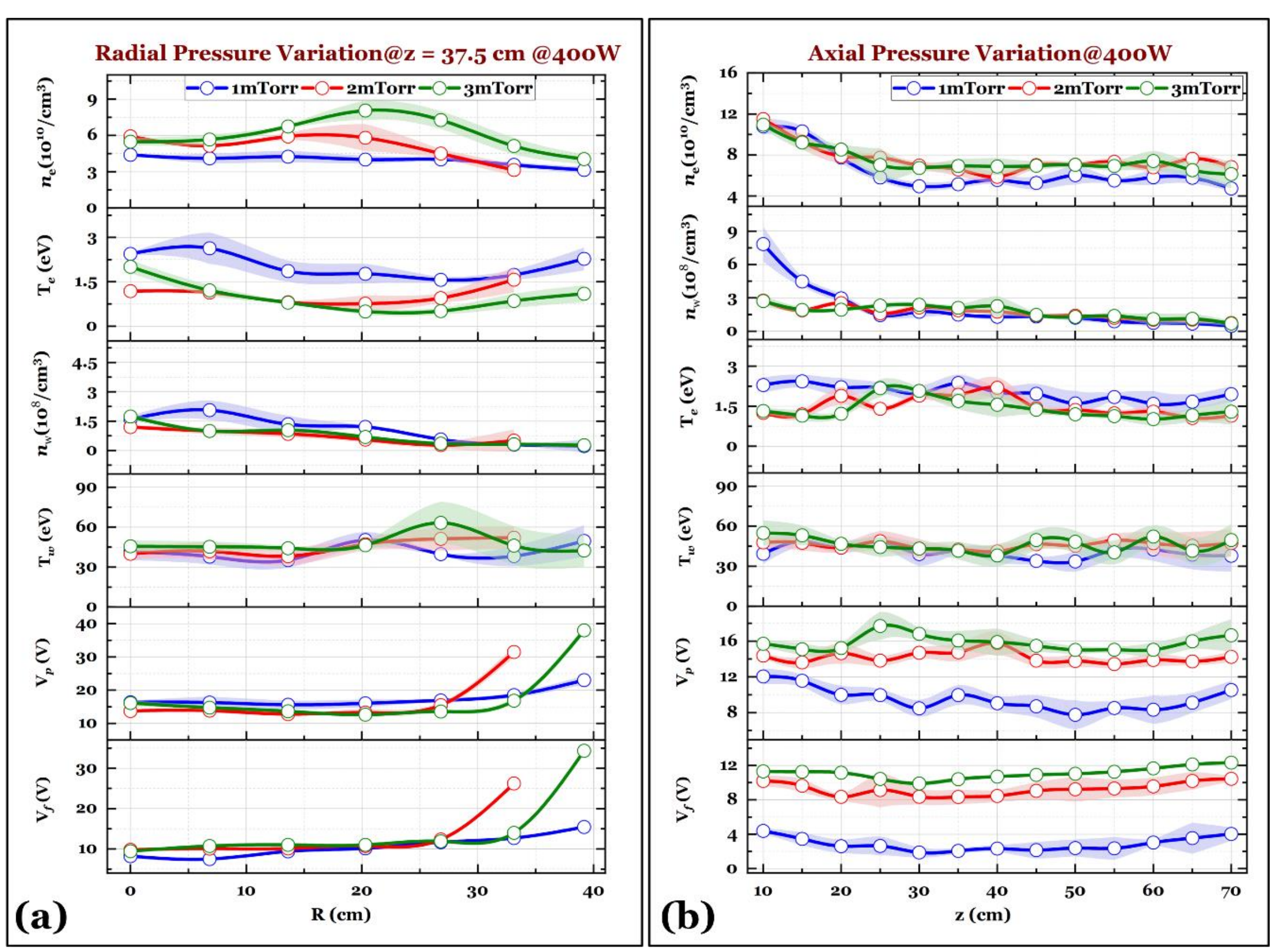

Figure 4. Radial variation of plasma parameters at z = 37.5 cm for configuration A (a), Axial variation of plasma parameters for configuration B (b).

### 3.1 Configuration A (CEPS source at the centre of the top dome)

Fig. 5 shows the $H^-$ IEDF measurement results in Configuration A (see Fig 1 (a)). It may be noted that in this configuration, because of the *obliquity of the endcap aperture* with respect to the direction of approach along the magnetic field, the IEDF count has to be amplified by the obliquity factor, $\cos^{-1}\theta \approx 1.39$ to adjust for the reduced collecting area. As can be seen from the figure, the measured IEDF exhibits an energy distribution with a shifted maximum, with the count peaking at an energy $E_p \approx$ 2 - 2.5 eV. The shifted distribution is characteristic of a beam and will be analysed later. *It is likely that the* $H^-$ *ions acquired this energy while sliding*

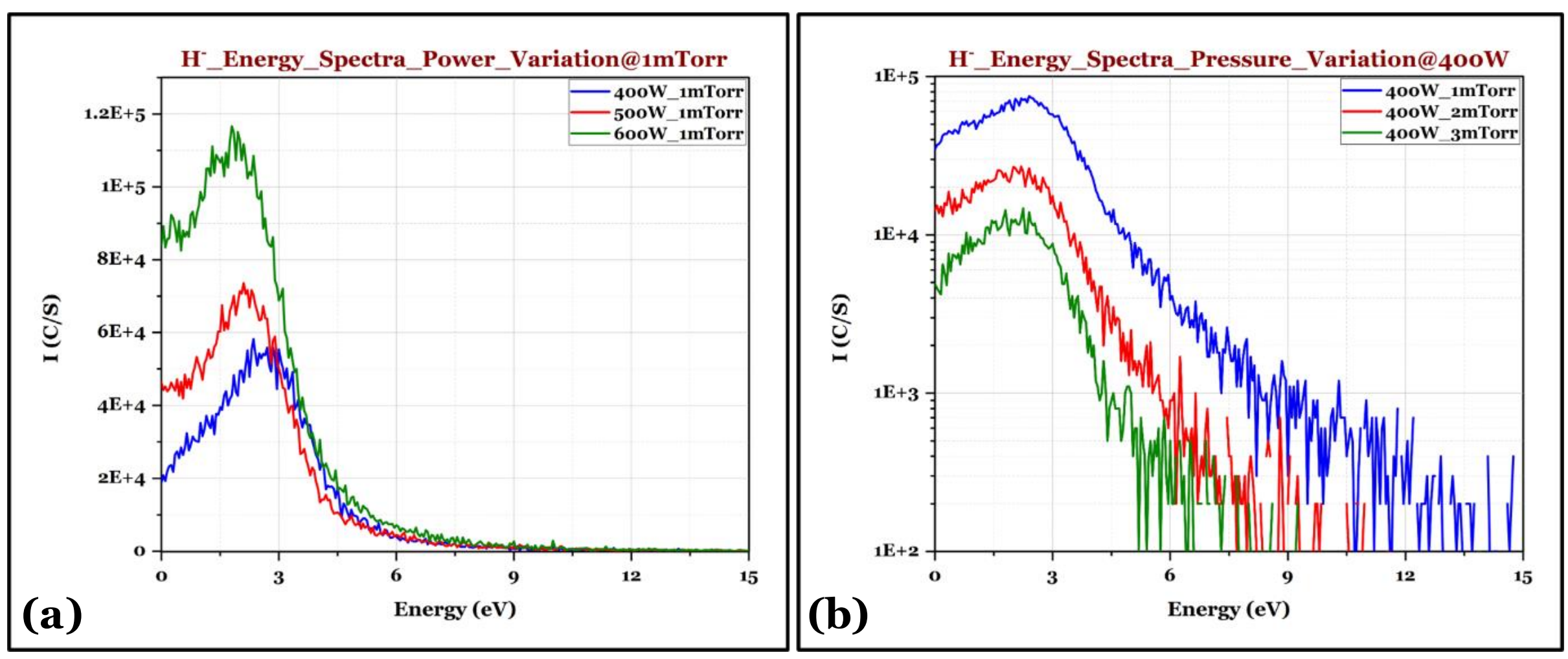

Figure 5. Power variation at 1 mTorr (a), Pressure variation at 400 W (b) of H- ion energy distribution function.

*down the magnetic field gradient along the field line leading to the PSMS probe (discussed later).* One finds that with an increase in microwave power from ≈ 400 W to ≈ 600 W at a constant pressure of ≈ 1 mTorr, the overall *peak counts* were enhanced monotonically from ≈ $5.8 \times 10^4\ C/s$ to ≈ $1.2 \times 10^5\ C/s$ without any major change in $E_p$. The increase in $H^-$ yields at higher microwave power is driven by improved power coupling within the plasma source (CEPS) region, resulting in an increase in plasma density there. Since the plasma flows out from the source region into the expansion chamber, one expects the plasma density in the downstream region to increase as well. It may be noted however, that as plasma flows into the expansion chamber, there is a steep fall in its density on account of strong expansion enforced by the diverging field lines. This expansion is accompanied by a cooling of the electrons possibly by adiabatic expansion of the electrons. In the downstream region therefore, these conditions create a favourable environment for dissociative electron attachment to the vibrationally excited $H_2$ molecules arriving from the source mouth ($z$ = 0), where they are formed by hot electrons with temperatures ≈ 15 – 20 eV). This issue will be again examined below.

Pressure variation studies at constant microwave power of ≈ 400 W show significant fall in $H^-$ counts with increasing gas pressure with $E_p$ remaining approximately the same. However, an increase in pressure implies that collisions would begin to dominate, leading to electron detachment of $H^-$ via interactions with the hydrogen molecules ($H^- + H_2(v) \rightarrow H + H_2(v') + e$) [23]. This causes the IEDF to scale down uniformly with the pressure across the entire energy spectrum while maintaining its characteristic shape. Although higher pressure enhances collisional cross-field diffusion ($D_\parallel \approx D_\perp$), it simultaneously increases the probability of $H^-$ destruction through collisional electron stripping via heavy particle collisions. Consequently, a significant fraction of $H^-$ ions are lost before reaching the detector, leading to a decrease in the measured flux without altering the intrinsic energy distribution. The transverse $H^-$ IEDF in Configuration A is characterized by a pronounced low-energy peak (representing a beam) and the absence of a high-energy tail. The radial distribution of the high energy $H^-$ ions is not known in this study. It will be seen from the results for Configuration B that energetic $H^-$ ions are definitely present on the source axis, although *their radial spread could be limited*. In the latter case, the energetic $H^-$ ions would continue along the axis, without being scattered away from the axis. A more detailed explanation in this regard is presented later.

While the diverging magnetic field (Fig. 1) guides the plasma into the expansion chamber, the outward curvature of the field lines simultaneously results in plasma flow towards the PSMS detector. In a diverging magnetic field, ion acceleration relies on the conversion of the energy of the magnetized electrons into the directed kinetic energy of the ions via ambipolar electric field. Electrons tend to diffuse downstream faster than the heavier ions and generates the ambipolar electric field, which pulls the ions and overall plasma forward to maintain plasma quasi-neutrality. If electrons are magnetized, plasma together with all the ions including the $H^-$ ions produced via DA process at higher planes ( $z > 10$ cm) follow the diverging field lines and reach the PSMS detector thereby possibly gaining the energy while sliding along a magnetic field gradient giving rise to the beam characteristics of the measured IEDFs. In low magnetic field zone while electrons follow natural motion, the ions move predominantly along the field lines, generated near the source chamber due to their inertia. This phenomenon can be explained by the parallel energy gain of ions as they move from a region of higher magnetic field to a lower magnetic field. Owing to the adiabatic conservation of the magnetic moment, the gradual decrease in magnetic field strength converts the ions' perpendicular energy into parallel kinetic energy along the magnetic field lines. This process is analogous to the magnetic mirror effect, where the reduction in magnetic field strength leads to a transfer of perpendicular thermal energy into parallel kinetic energy while conserving the magnetic moment. A detailed theoretical derivation of this phenomena is provided in Appendix A.

Fig. 6 shows the recorded $H^-$ energy distribution at 400 W – 1 mTorr fitted analytically with a combination of Maxwellian and Normal distribution. Normalized individual distributions have also been plotted for comparison. The functional form of the fitting is represented in equation (1).

$$f(E) = a_{MB} f_1 + a_N f_2 \tag{1}$$

Where,

$$f_1 = \left(\frac{2}{\sqrt{\pi}}\right)\left(\frac{\sqrt{E}}{T_i^{\frac{3}{2}}}\right) exp\left(-\frac{E}{T_i}\right)$$

$$f_2 = \frac{1}{S\sqrt{2\pi}}\, exp\left(-\frac{(E - E_0)^2}{2S^2}\right)$$

$a_{MB}$ = Weight factor for Maxwell-Boltzmann Distribution

$a_N$ = Weight factor for Normal Distribution

$E$ = Ion Energy

$T_i$ = Ion Temperature

$E_0$ = Peak Energy of Normal Distribution

$S$ = Full width at half maximum (FWHM) of Normal Distribution

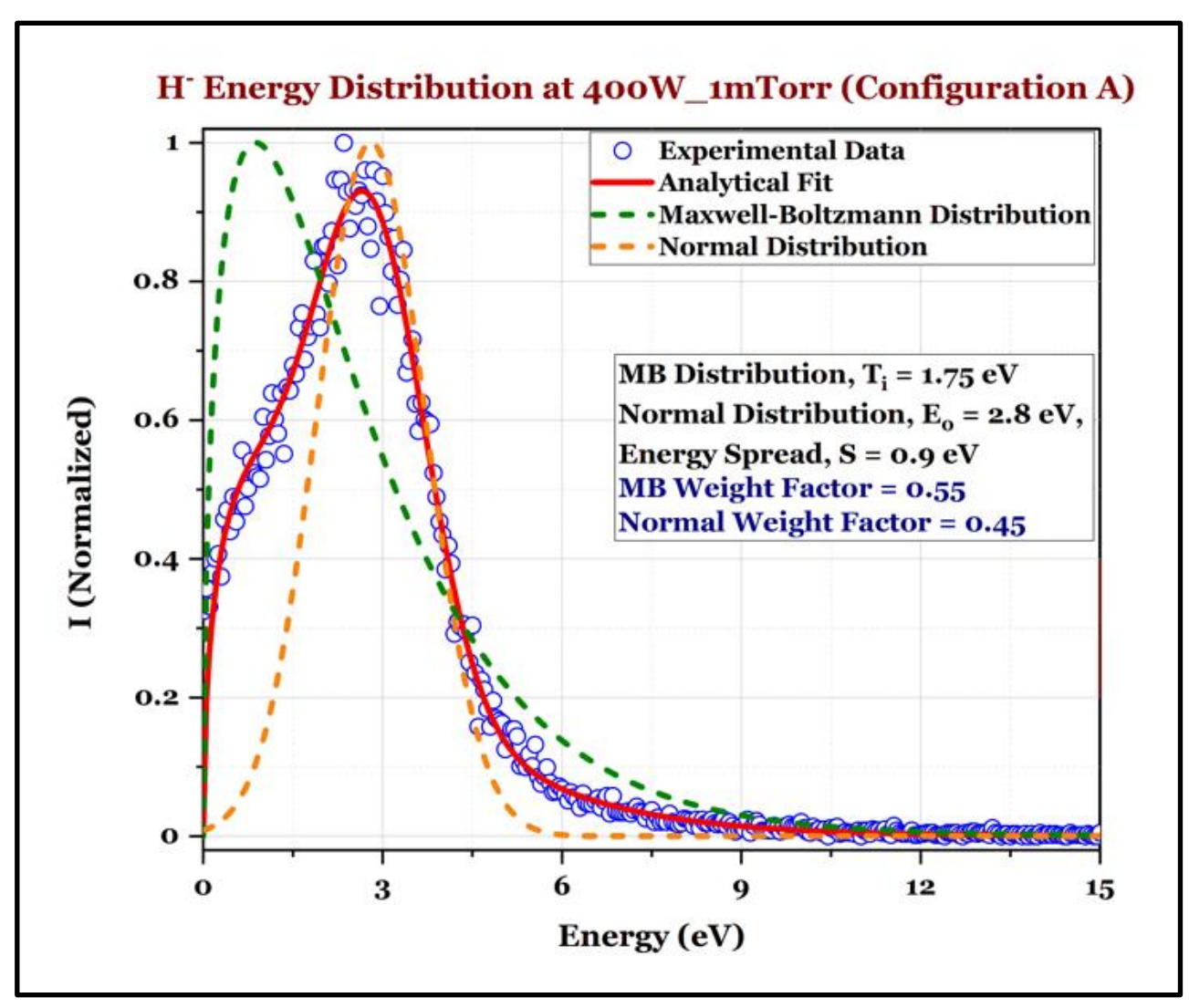

Figure 6. Analytical fitting of the experimentally measured H- energy distribution

Using the best fit to the measured IEDF, the thermalized Maxwellian ion population with a temperature of ≈1.75 eV accounts for approximately 55 % of the total distribution ($a_{MB}$ = 0.55). This component is attributed to $H^-$ ions produced via dissociative attachment (DA) somewhere in the diverging magnetic field region. In addition, 45 % of contribution arises from a Normal distribution centred at ≈ 2.8 eV with an energy spread of ≈ 0.9 eV, indicating the presence of an ion beam population likely associated with the acceleration of $H^-$ ions along the diverging magnetic field line as already discussed.

### 3.2 Configuration B (CEPS source at the sidewall facing MBMS PSMS probe)

The mass spectrometer measurements were repeated in Configuration B as shown in Fig 1(b). This allows one to determine the mass / energy spectra along the direction of the plasma flow and compare the results with earlier measurements along the transverse direction. Fig. 7 shows the power and pressure variation IEDF data for $H^-$ ions in configuration B. The ion energy distribution is characterized by a shifted normal distribution at low-energies peaking at $E_p \approx 2$ eV, accompanied by a high-energy tail extending up to ≈ 20 eV at 1 mTorr. With increasing microwave power, the detected $H^-$ ion counts decrease, with a more pronounced reduction observed at higher energies. No significant change in $E_p$, the peak energy was observed with increasing microwave power. Furthermore, no noticeable change in the distribution function was observed.

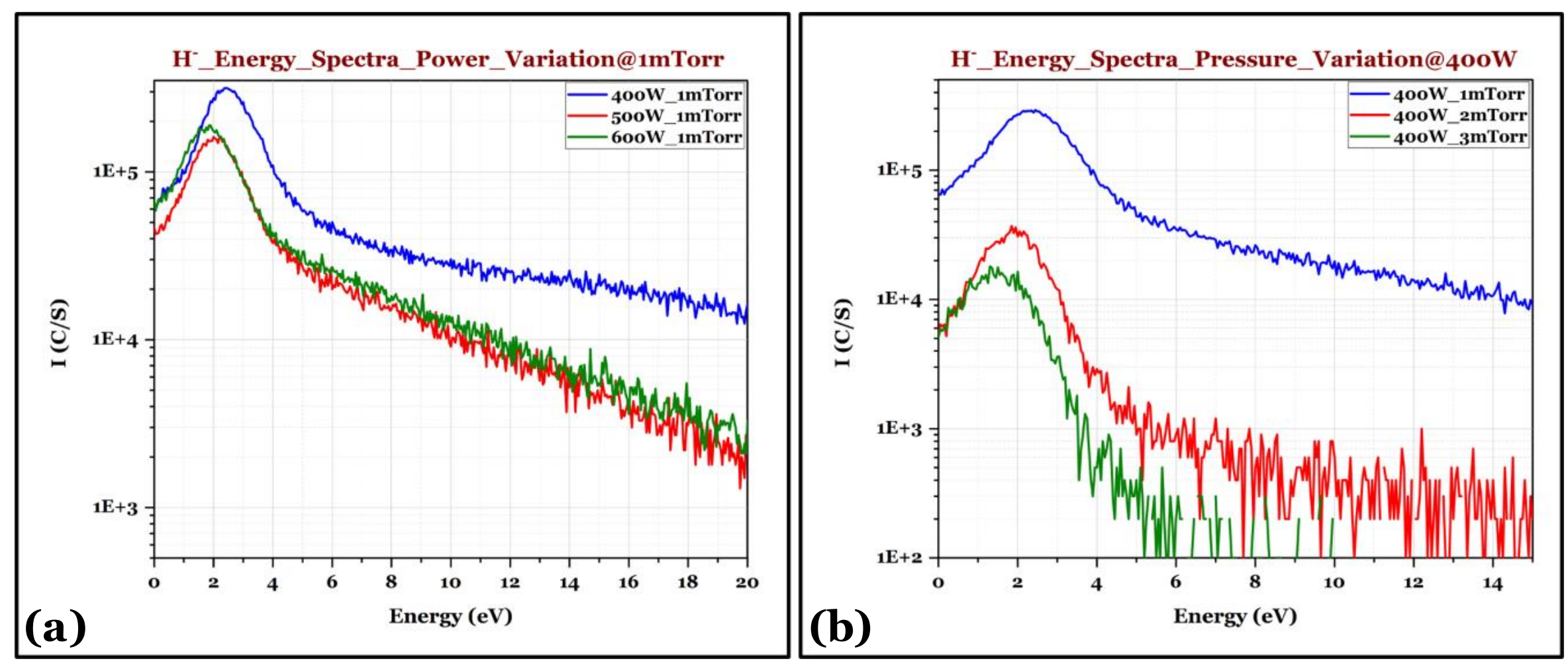

Figure 7. Power variation at 1 mTorr (a), Pressure variation at 400 W (b) of H- ion energy distribution function

The abnormal *decrease* of the $H^-$ count with microwave power (at ≈ 1 mTorr) in Configuration B may be understood by considering the following scenario. Since the formation of the $H^-$ ions depends sensitively on the requirement that $T_e$ should be low (≲ 1.5 eV), even a slight *increase in the on-axis value of $T_e$ with microwave power could bring about a corresponding decrease in the population of the low-$T_e$ electrons, which in turn could result in a fall in $H^-$ formation and its subsequent count at the MBMS*. This is particularly true for Configuration B where the PSMS probe is located on the axis, in direct line of sight of the source. On the other hand, the off-axis behaviour of $T_e$ ($r \approx 30$ cm) could remain unaffected (or show a slight fall) with possible increase of $n_e$. In such a case, it would be possible the $H^-$ count would increase with power as was seen for Configuration A. The high energy tail in the $H^-$ IEDF in Fig. 7 is possibly due to an increasing plasma potential (accelerating $H^-$ ions). The reduction of the high energy population with microwave power is possibly due to a flattening of the profile with distance along the axis. These arguments seem plausible although the limited LP data available seem to support these statements. With increasing pressure, the detected $H^-$ ion counts decrease and the nature of the distribution changes with the high-energy tail progressively disappearing. This behaviour follows from the combined effects of enhanced collisionality and energy-dependent loss mechanisms discussed earlier.

Fig. 8 shows the recorded $H^-$ energy distribution at ≈ 500 W - 1 mTorr analytically fitted with a combination of Maxwellian and Normal distribution. Decomposition of the measured IEDF into two components reveals that the Maxwellian ion population with temperature ~ 4 eV contributes approximately 54 % of the total distribution which contributes to the high energy tail part of the distribution. In addition to it, 46 % contribution arises from a shifted Normal distribution, peak centred at ≈ 2.0 eV with an energy spread of ≈ 0.84 eV showing the presence of an ion beam population associated with the $H^-$ ions reaching the detector from the upper section of the expansion chamber (z > 10 cm) accelerated along the diverging magnetic field. Compared to configuration A, configuration B exhibits an pronounced high energy tail thermal distribution. A detailed comparison of the results for the Configuration A and B is presented in the next section.

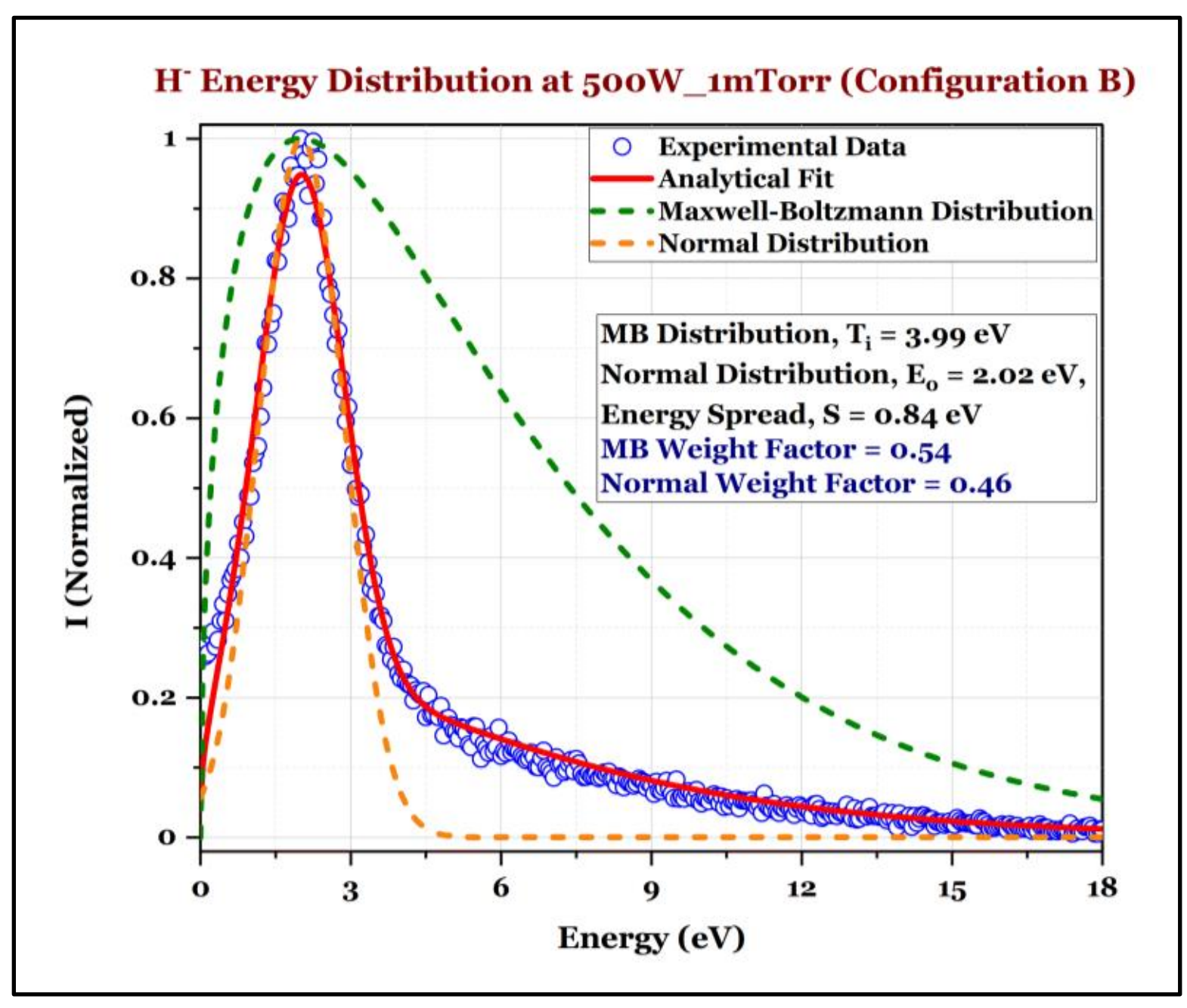

Figure 8. Analytical fitting of the experimentally measured $H^-$ energy distribution

# 4. Analysis and Discussion

## 4.1 Comparison of Configuration A and Configuration B

The comparison of the $H^-$ IEDFs for the two configurations at two different operating pressures is presented in Fig. 9 where the effect of both the source configuration and gas pressure on $H^-$ IEDF can be seen. The most significant difference is in the higher energy $H^-$ count in Configuration B. At ≈ 1 mTorr the difference is seen across all energies, whereas at ≈ 3 mTorr, the difference is mainly at low energies.

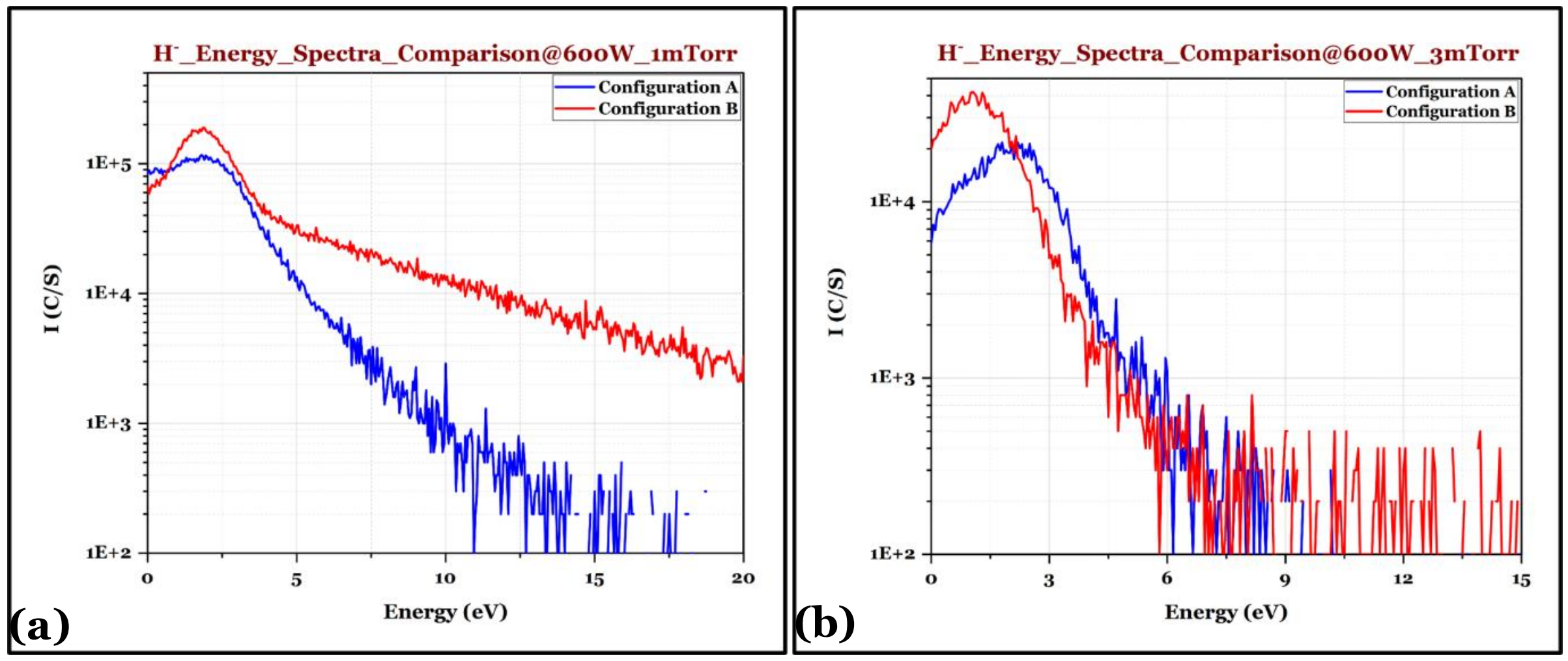

Figure 9. Comparison of negative ion energy distribution for the two configurations at 600 W - 1 mTorr (a) & 600 W - 3 mTorr (b)

One can try to understand this difference by considering the ease with which the $H^-$ ions move along and transverse to the magnetic field (in configurations A and B) to reach the PSMS probe. For instance, the observed difference between the measured $H^-$ IEDFs at ≈ 1 mTorr indicates that energetic $H^-$ ions are transported more efficiently in Configuration B than in Configuration A. To understand the reason for this one needs to consider the parallel and perpendicular diffusion coefficients of the $H^-$ ions. Figures. 10a and 10b show the variation of the parallel (($D_{\parallel}$) and perpendicular ($D_{\perp}$) diffusion coefficients with $H^-$ ion energy for Configurations A and B at ≈ 1 mTorr and ≈ 3 mTorr, respectively, calculated at the PBMS probe endcap. The diffusion coefficients are defined as [24]:

$$D_{\parallel} = \frac{KT_i}{m\nu_m} \tag{2}$$

$$D_{\perp} = \frac{D_{\parallel}}{1+\left(\frac{\omega_c}{\nu_m}\right)^2} \quad (3)$$

Here,

$$D_{\parallel} = Parallel\ diffusion\ coefficient$$

$$D_{\perp} = Perpendicular\ diffusion\ coefficient$$

$$K = Boltzmann's\ Constant$$

$$T_i = Ion\ Temperature$$

$$m = Mass\ of\ H^{-} ions$$

$$\nu_m = Momentum\ transfer\ collision\ frequency$$

$$\omega_c = Ion\ cyclotron\ frequency$$

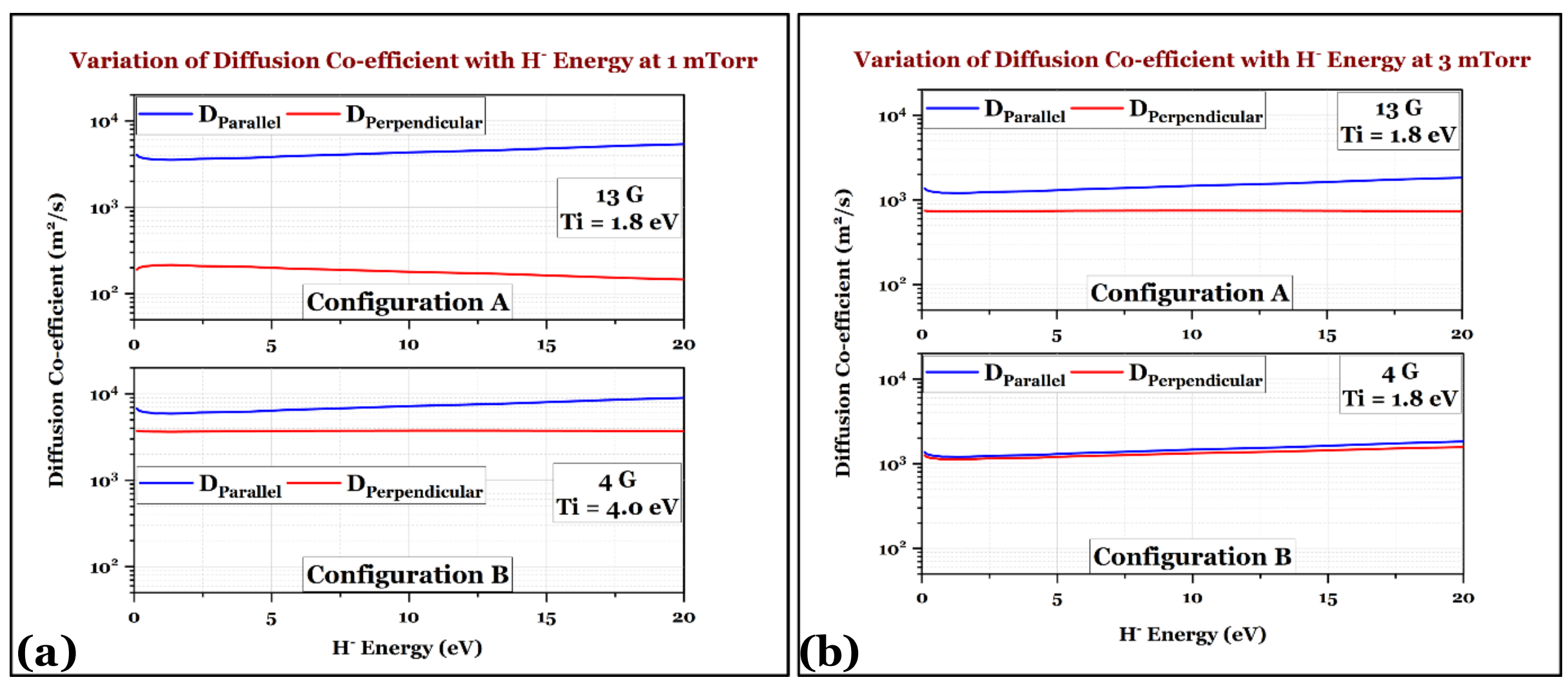

Figure 10. Variation of diffusion co-efficient with $H^-$ energy at (a)1mTorr and (b) 3mTorr

It should be noted that the calculated diffusion coefficients show very little variation with $H^-$ energy and indicate no intrinsic dependence on the orientation of the PSMS probe relative to the magnetic field. Instead, the observed differences between Configurations A and B arise primarily from the local magnetic field strength at the detector location. The magnetic field at the PSMS position is significantly stronger in Configuration A than in Configuration B. Consequently, for Configuration A at ≈ 1 mTorr, $\left(\left(\frac{\omega_c}{\nu_m}\right)^2 \gg 1\right)$ implying $D_{\parallel} >> D_{\perp}$, and suppression of cross-field transport. Under these conditions, $H^-$ ions are transported predominantly along magnetic field lines, limiting the number of ions that can reach the detector. In contrast, the magnetic field at the PSMS location in Configuration B is weaker $\left(\left(\frac{\omega_c}{\nu_m}\right)^2 \ll 1\right)$ and $D_{\parallel} \approx D_{\perp}$, implying isotropic diffusion. This allows the $H^-$ ions to reach the detector from off-axis locations as well, giving a higher $H^-$ count. At ≈ 3 mTorr, the collision frequency increases significantly, reducing the influence of magnetic-field-controlled transport. Under these conditions, diffusion becomes isotropic. Consequently, the distinction between the IEDFs measured in Configurations A and B is reduced and the high-energy component observed in Configuration B at ≈ 1 mTorr is no longer present.

In addition to modifying transport, the increase in pressure also enhances electron-detachment (stripping) losses of $H^-$ ions during their propagation from the expansion region to the detector. The higher collision frequency increases the probability of electron detachment reactions, reducing the number of $H^-$ ions that survive the transport path. Consequently, the measured $H^-$ ion counts decrease for both configurations with increasing pressure. Despite the reduction in ion flux, the peak position of the IEDF remains nearly unchanged for both configurations. This observation suggests that the most probable energy of the detected $H^-$ ions is governed primarily by the volume production mechanism responsible for their formation. While transport processes and collisional losses influence the number of ions reaching the detector, they do not significantly alter the characteristic energy associated with the dominant $H^-$ production process.

## 4.2 Relative yield percentage of positive ions

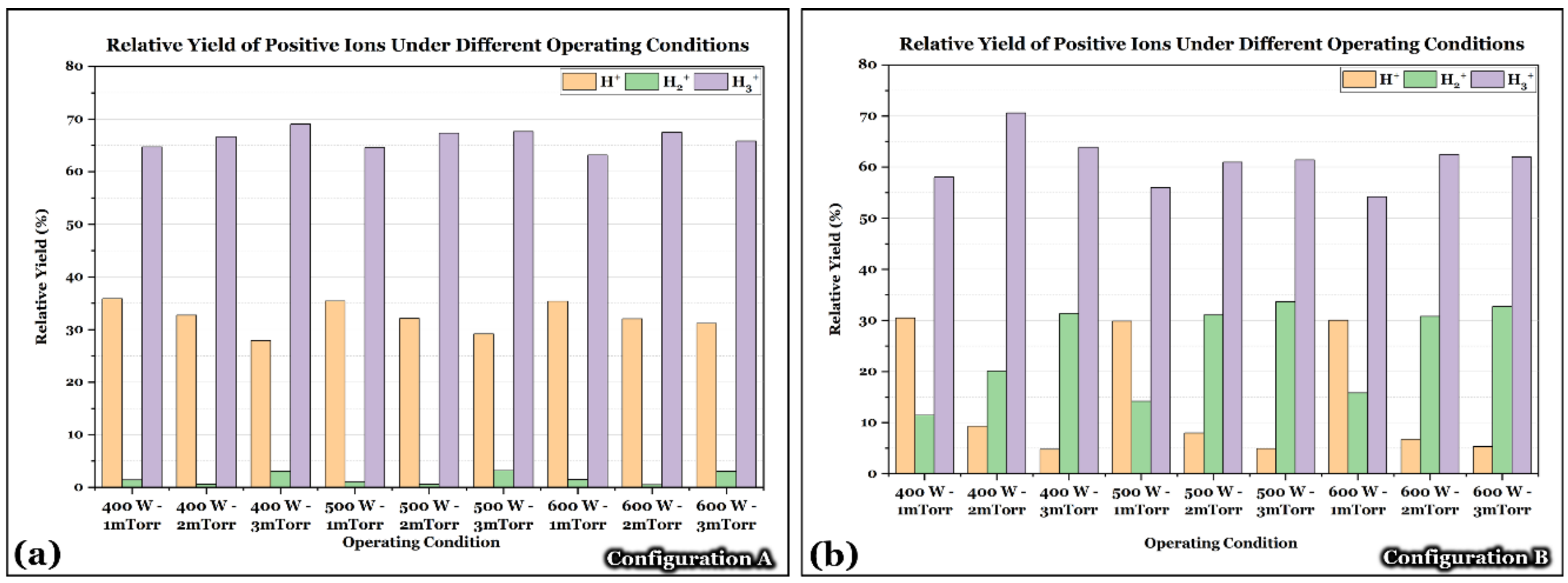

Figure 11. Relative percentage yield of positive ions for the two configurations: (a) Configuration A, (b) Configuration B

Measurements related to the hydrogen positive ions were also undertaken with the two configurations. The relative percentage yield of the positive ions for the two studied configurations are presented in Fig. 11. Across all the experimental conditions studied, $H_3^+$ remains the dominant ion species for both the configurations A and B. For Configuration A, $H_2^+$ ions are the least abundant across all power and pressure ranges studied. However, the same is not valid for Configuration B. At 1 mTorr, the $H_2^+$ ions are the least abundant species, but the trend shifts at higher pressures (2 - 3 mTorr) where $H^+$ ions become the least abundant species. This indicates pressure dependent production and transport mechanisms that modifies the detection probabilities of ions in two different measurement directions. The dominant relative yield of $H_3^+$ ions in both the configurations suggests that a pathway involving $H_3^+$ may provide an additional channel for $H^-$ formation through the process $(e + H_3^+ \rightarrow H_2^+ + H^-)$ [25], alongside the primary dissociative electron attachment (DEA) mechanism. A more detailed insight about the underlying plasma chemistry and transport process responsible for the same is under investigation.

### 4.3 Density estimation of positive and negative hydrogen ions

An approximate estimation of the positive and negative ion densities can be obtained by combining the relative yields of ions measured using the MBMS with the ion saturation current measured by the Langmuir probe. The details of the calculation procedure are provided in Appendix A. Briefly, the relative MBMS counts corresponding to the peak energy of $H^+, H_2^+ \; and \; H_3^+$ were used together with the Bohm flux relation to estimate the individual positive ion densities, while the measured H⁻ counts were used to estimate the negative ion density utilizing the conversion factor for positive ions. Although the method relies on several simplifying assumptions, it provides a reasonable estimate of the density of ion species present at the measurement site. The analysis for ≈ 500W, ≈ 2mTorr in Configuration B shows that the positive ion population is dominated by $H_3^+$, followed by $H_2^+$, while $H^+$ constitutes the smallest fraction. The estimated densities were:

$$n_{H^+} \approx 9.6 \times 10^9 \; cm^{-3}$$

$$n_{H_2^+} \approx 1.7 \times 10^{10} \; cm^{-3}$$

$$\text{and } n_{H_3^+} \approx 4.3 \times 10^{10} \; cm^{-3}$$

Yielding a total positive ion density of approximately:

$$\Rightarrow n_i \approx n_e = (n_{H^+}) + (n_{H_2^+}) + (n_{H_3^+}) = 7.0 \times 10^{10} \; cm^{-3}$$

The dominance of $H_3^+$ for our operating power-pressure regime is consistent with the low electron temperature (≈ 2 eV) measured in the downstream plasma, where ion–molecule conversion reactions efficiently convert $H_2^+ \; into \; H_3^+$ via the reaction ($H_2 + H_2^+ \rightarrow H_3^+ + H$) as reported in ref. [26].

Using the same approach, the H⁻ ion density was estimated to be:

$$n_{H^-} \approx 3.9 \times 10^8 \; cm^{-3}$$

This corresponds to a plasma electronegativity, $n_{H^-} / n_e \approx 0.55\%$, where $n_e$ is the electron density as determined by the quasineutrality condition stated above. The relatively low electronegativity indicates that the plasma remains predominantly electropositive under the present operating conditions. However, it should be noted that this value represents the H⁻ density measured at the PSMS probe location and should not be interpreted as the maximum H⁻ density produced by the source. In Configuration B, the measurement location is approximately 80 cm downstream from the source exit, where the plasma has already undergone substantial expansion and the negative ions have experienced transport as well as electron-detachment losses. Consequently, the calculated density reflects only the surviving H⁻ population reaching the detector transported through the expanding plasma volume.

*One can estimate the H⁻ density near their formation zone inside the expansion chamber by accounting for transport losses due to scattering and destruction of the negative ions, between their production region and the diagnostic location. To obtain an estimate of the H⁻ density produced upstream (z ≳ 10 cm), one may calculate an effective mean free path for the scattering*

*and destruction of H⁻ ions through momentum transfer collisions with hydrogen molecules, electron detachment collisions with electrons, atomic hydrogen, and molecular hydrogen, and mutual neutralization with positive ions [23].* The resulting effective mean free path was found to be:

$$\lambda_{eff} \approx 12.37\ cm$$

Based on the earlier plasma characterizations [22], the dominant $H^-$ production zone is expected to lie approximately at $z \approx 10$ - 30 cm from the source mouth. However, as the ions travel, their density falls with distance. The scattering and loss processes (stated above) arise as a result of *collisions*, specific to each type of encounter.

*Collisions are random events that obey the exponential probability distribution*. This implies that the probability that an ion starting out from the location $z$, will reach a location $z_0$, *without suffering any collision* is $\propto \exp[-\frac{(z_0-z)}{\lambda}]$, where $\lambda$ is the mean free path.

Using $\lambda_{eff}$ for the *effective mean free path* for $H^-$ ions and letting $n_{H^-}(z)$ and $n_{H^-}(z_0)$ be the $H^-$ densities at the *production site z* and the *detection site* $z_0$, respectively, obtains

$$n_{H^-}(z_0) = n_{H^-}(z) \exp\left(-\frac{(z_0 - z)}{\lambda_{eff}}\right) \tag{4}$$

It may be noted that though the production site $z$ lies *upstream* of the detection site $z_0$, it is $n_{H^-}(z_0)$ *that is known from the measurements*. Thus (4) may be inverted to estimate the density $n_{H^-}(z)$ at the production site $z$.

$$n_{H^-}(z) = n_{H^-}(z_0) \exp\left(+\frac{(z_0 - z)}{\lambda_{eff}}\right) \tag{5}$$

Assuming the $H^-$ ions are produced uniformly in the region $z = 10 - 30$ cm, one may calculate the average density, $\langle n_{H^-}(z)\rangle$ over the production zone, noting that $z_0 \approx 80$ cm and $\lambda_{\text{eff}} \approx 12.37$ cm. $\langle n_{H^-}(z)\rangle$ is given by

$$\langle n_{H^-}(z)\rangle = \frac{n_{H^-}(z_0)}{z_2 - z_1} \int_{z_1}^{z_2} \exp\left(+\frac{(80 - z)}{\lambda_{eff}}\right) dz \tag{6}$$

Solving (6) gives

$$\langle n_{H^-}(z)\rangle \approx 142\ n_{H^-}(z_0) \approx 5.5 \times 10^{10}\ \text{cm}^{-3}$$

Here $n_{H^-}(z_0) \approx 3.9 \times 10^8\ \text{cm}^{-3}$ was used.

Although this estimate is subject to uncertainties arising from the assumed production profile and the spatial variation of plasma parameters, it nevertheless provides a useful indication of the $H^-$ generation capability of the source. The estimated density is significant considering that these results were obtained at a relatively modest microwave power of only ≈ 500 W. Notably, the estimated $H^-$ density is comparable to the values reported for filament sources [27,28] and RF- driven Helicon sources [29,30] of smaller sizes operating at kilowatt-level powers that demonstrates the efficiency of ECR-based CEPS source.

### 4.4 Working of the CEPS and its role in enhancement of $H^-$ density:

The uniqueness of the CEPS lies in its ability to produce energetic electrons close to the source exit and low-temperature downstream plasma that favours volume production of $H^-$ ions. Furthermore, the CEPS sustains a uniform hydrogen plasma over a large volume with an effective diameter of approximately 60 cm (due to the fanning out of its magnetic field). The observation of substantial $H^-$ densities after transport over such a long distance demonstrates the capability of the source to generate and sustain negative hydrogen ions over extended plasma volumes, highlighting the effectiveness of the CEPS for large-area negative ion production for fusion applications.

The significantly high $H^-$ density estimated in the downstream region ( ~ 10 – 30 cm from the source mouth) may be attributed to the operational efficiency and unique configuration of the CEPS. It is worth recalling that the volume production rate of $H^-$ ions is proportional to both the density of low-temperature electrons $(T_e \leq 1.5\ eV)$ and the population of vibrationally excited hydrogen molecules, $H_2(v^*)$. The latter are expected to be generated efficiently within the plasma source section (PSS), extending from the resonance region to the vicinity of the source mouth, where the electron temperature remains relatively high $(T_e > 20\ eV)$, as confirmed by previous experiments involving CEPS with hydrogen plasma [22].

High-energy electrons generated in the ECR zone efficiently collide with neutral $H_2$ molecules, exciting them to higher vibrational levels. This process can be understood in greater detail by considering the plasma dynamics within the CEPS. The magnetic field in the ECR zone within the CEPS has the structure of a 3D magnetic mirror (see Fig. 3a) that affords multiple encounters of the mirror-trapped electrons with the resonant microwaves, which energizes them thereby enabling them to ionize the gas molecules. The process continues till collisions deconfine the electrons, allowing them to escape into the adjacent plasma source section (PSS) towards the exit. The hot electrons escaping initially leave behind positively charged ions that establish a high plasma potential within the source region. This potential produces an ambipolar electric field that inhibits the rapid escape of electrons while simultaneously driving the colder ions outward. Under steady-state conditions, both species eventually flow together toward the source mouth to preserve quasineutrality.

As the plasma propagates toward the PSS exit, the hot electrons initiate a second ionization zone (SIZ) near the source mouth, which facilitates the formation of a double layer (DL). The SIZ enhances the plasma density (accompanied by a fall in the electron temperature), whereas the DL accelerates the ions and produces a high-energy ion beam [31]. The sharp reduction in $T_e$ across the SIZ is followed by a more gradual decrease over a distance of approximately ≈ 10 cm, with $T_e$ eventually reaching approximately ~ 1 eV.

Within the PSS, electrons and ions propagate with a common ambipolar flow velocity that is much lower than the electron thermal velocity. The reduction in electron velocity is caused by the combined effects of the ambipolar electric field and frequent elastic collisions with neutral particles. Under steady-state conditions, the electrons are therefore slowed to approximately the flow velocity, $v$ of the ions. In principle, $v$ can be determined by solving the ion momentum equation in the paraxial region between the ECR zone and the SIZ, including the effects of the steady-state ambipolar electric field, as well as friction associated with elastic and charge-exchange collisions with neutrals.

Without considering the details of such a solution, however, an order-of-magnitude estimate of the flow velocity can be obtained from the potential drop across the SIZ. This potential drop can reach up to approximately ≈ 80 V, as reported in previous experiments with hydrogen plasmas in CEPS [18]. An $H^{+}$ ion accelerated through such a potential difference would acquire a velocity of $\approx\ 1.24\ \times\ 10^{5}\ m/s$. This value may, therefore, be regarded as the characteristic flow velocity of both ions and electrons within the PSS.

On the other hand, the collision frequency (ν), for vibrational excitation of $H_2$ by energetic electrons is approximately ≈ 2.5 × $10^{6}$ $s^{-1}$ over the present pressure range. The corresponding excitation mean free path $\lambda = v/\nu$, is therefore approximately $\approx\ 5\ cm$. This length is substantially shorter than the combined distance of approximately 12 cm spanning the region between the ECR zone and the SIZ and the region over which $T_e$ subsequently decreases gradually to approximately 1 eV. Consequently, energetic electrons have a high probability of undergoing vibrational excitation of $H_2$ within this region, thereby producing a substantial population of vibrationally excited molecules, $H_2(v^{*})$.

Once transported into the downstream low-temperature region, these vibrationally excited molecules, which are unaffected by the magnetic field, can participate in the production of $H^{-}$ ions through dissociative attachment (DA). It is also noteworthy that the ionization mean free path is of the same order as the excitation mean free path estimated above, thereby providing a plausible explanation for the initiation of the SIZ. The enhancement of the electron density within the SIZ consequently supplies an adequate population of electrons downstream to sustain the DA reaction and, hence, the observed high $H^-$ density.

## 5. Summary and Inference

The present study provides a detailed characterization of $H^-$ IEDFs in a microwave-driven ECR hydrogen plasma source under different operating conditions and measurement configurations. Earlier plasma characterization established favourable downstream conditions for volume $H^-$ production, characterized by low electron temperatures and sufficiently high plasma densities. Building on these observations, $H^-$ IEDFs were measured using a Hiden Analytical HPR-60 MBMS in two configurations: transverse to the plasma expansion direction (Configuration A) and along the plasma expansion direction (Configuration B). In Configuration A, $H^-$ ion counts increased with microwave power, consistent with enhanced plasma density while maintaining conditions favourable for dissociative electron attachment. The measured $H^-$ counts decreased with increasing pressure, primarily due to enhanced collisional electron-detachment losses during transport from the production region to the detector. In Configuration B, the $H^-$ IEDFs exhibited a pronounced high-energy component at 1 mTorr that progressively diminished with increasing pressure. Unlike Configuration A, the $H^-$ counts decreased with increasing microwave power, likely due to a rise in electron temperature within the $H^-$ formation region, reducing the population of low-energy electrons required for efficient $H^-$ production. As expected, increasing pressure also resulted in lower $H^-$ counts because of enhanced collisional losses.

An approximate estimation of the ion densities was performed by combining the relative ion yields obtained from MBMS measurements with the ion saturation current measured using a Langmuir probe. The positive ion population was found to be dominated by ${H_3}^+$ ions, followed by ${H_2}^+$ and $H^+$ ions, yielding a total positive ion density of $\approx 7.0 \times 10^{10}\ cm^{-3}$ at the measurement location under operating conditions of ≈ 500 W and ≈ 2 mTorr. The corresponding $H^-$ density was estimated to be $\approx 4.0 \times 10^{8}\ cm^{-3}$. Using the effective mean free path of ≈ 12.37 cm obtained by accounting for the cumulative effects of collisional scattering, electron detachment and mutual neutralization losses during transport, a simple calculation yields the $H^-$ density to be $\approx 5.5 \times 10^{10}\ \mathrm{cm}^{-3}$ in the production region $z \approx (10-30)\ cm$. Despite the relatively modest microwave power of ≈ 500 W, the inferred $H^-$ density is comparable to values reported for several volume-mode $H^-$ sources operating at significantly higher powers and with smaller plasma volumes. This high $H^-$ density is attributed to the favourable plasma conditions established by the CEPS configuration, where efficient ECR heating promotes the formation of $H_2(v)$ molecules, while downstream electron cooling provides the low-energy electrons required for dissociative attachment. The coexistence of these conditions over an extended downstream region enables efficient volume production of $H^-$ ions and demonstrates the potential of the CEPS source for large-area negative ion source applications.

Overall, the work demonstrates that the measured $H^-$ ion dynamics are governed by a combined influence of plasma production, magnetic-field-dependent transport, and collisional loss processes. The ability of the source to generate and sustain a relatively high density of $H^-$ ions over large plasma volumes at relatively low microwave power, together with the observed transport characteristics, provides valuable insights for tailoring plasma conditions and source geometry to maximize $H^-$ production, offering critical guidance for the development of energy-efficient next-generation large-area negative ion sources for fusion applications.

## Appendix A

## Parallel Energy Gain of $H^-$ Ions in a Diverging Magnetic Field

This process can be understood by examining Fig. 12. Let the field line passing through P1 enter the PSMS probe at P2. Consider $H^-$ ions that are formed at P1. These ions will flow along the field lines passing through P1 to reach P2. Conservation of total kinetic energy requires:

$$E_{\parallel 1} + E_{\perp 1} = E_{\parallel 2} + E_{\perp 2} \tag{1}$$

The magnetic moment, μ can be expressed in terms of energy $E_\perp$ and magnetic field $B$ as:

$$\mu = \frac{mv_\perp^2}{2B} = \frac{E_\perp}{B} \tag{2}$$

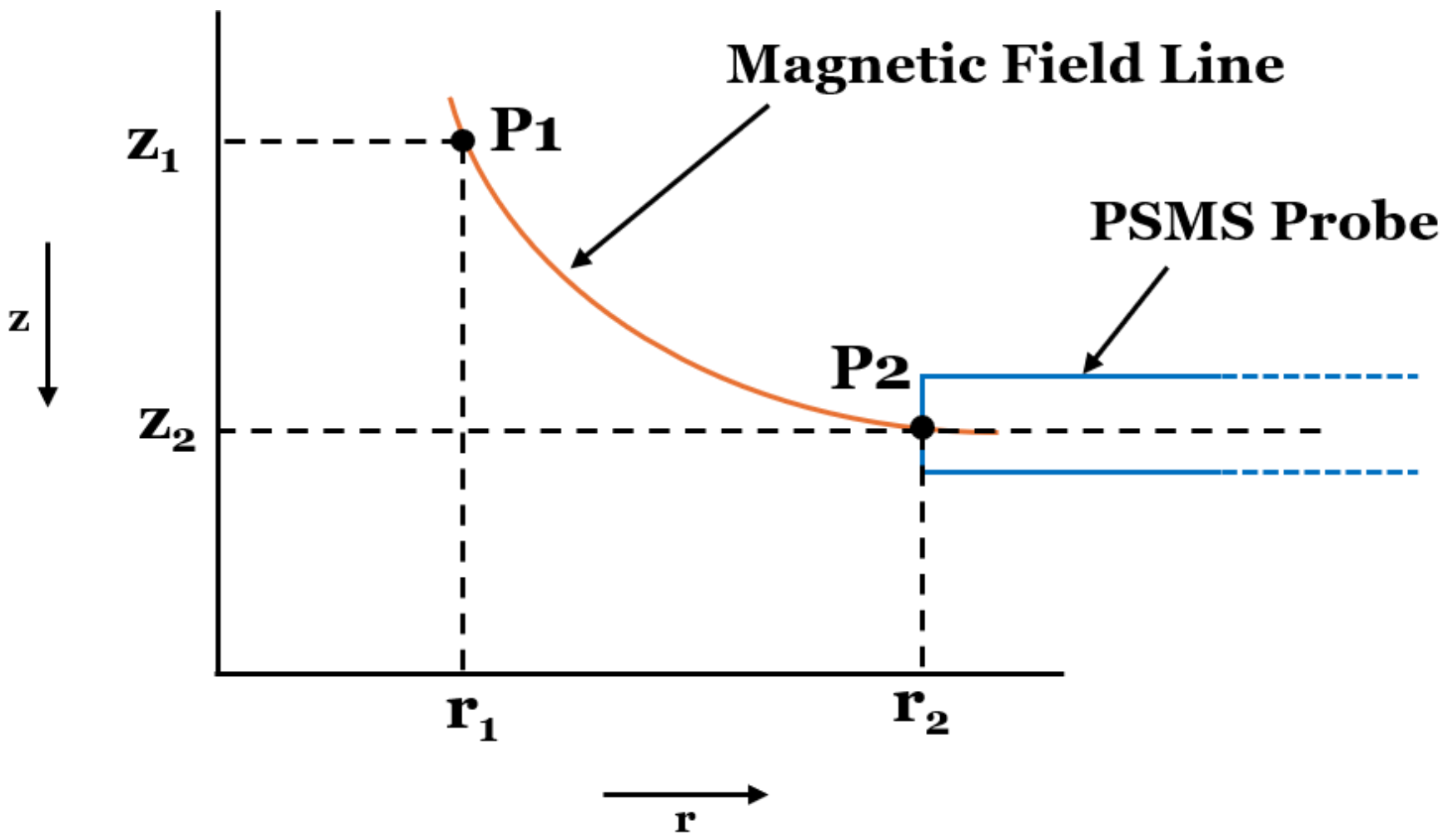

Figure 12. Acceleration of H⁻ ions along diverging magnetic field

Averaging (1) and (2) over the perpendicular kinetic energy gives

$$\langle\mu\rangle_v = \langle\frac{E_\perp}{B}\rangle_v = \frac{T_\perp}{B} \tag{3}$$

After averaging, Eq. (1) becomes

$$E_{\parallel 1} + T_{\perp 1} = E_{\parallel 2} + T_{\perp 2} \tag{4}$$

By the adiabatic invariance of magnetic moment along the field line,

$$\frac{T_{\perp 1}}{B_1} = \frac{T_{\perp 2}}{B_2} \tag{5}$$

$$\Rightarrow T_{\perp 2} = T_{\perp 1}\frac{B_2}{B_1} \tag{6}$$

Putting (6) in (4) we get:

$$T_{\perp 1}\left(1 - \frac{B_2}{B_1}\right) = E_{\parallel 2} - E_{\parallel 1} \tag{7}$$

Assuming that $H^-$ ions formed at P1 have negligible parallel energy, $E_{\parallel 1} \approx 0$, gives:

$$T_{\perp 1} = \frac{E_{\parallel 2}}{1 - \frac{B_2}{B_1}} \tag{8}$$

Using equation (8) one may determine the perpendicular energy of the $H^-$ ions in order for them to have the energy $E_{\parallel 2}$ measured at P2 by the PSMS probe provided the magnetic field value are known. One needs to locate P1 on the field line joining P2 and P1, so that $T_{\perp 1}$ has a reasonable value. Taking $E_{\parallel 2} \approx 2.5$ eV from the measured MBMS data and $B_2 = 12.6$ G one finds, *moving along the field line from P2*, that the point $z_1 \approx 10$ cm, and $r_1 \approx 8$ cm, has a value for $B_1 \approx 194.3$ G. This point may be associated with P1, since it gives from (8), $T_{\perp 1} = 2.65$ eV, which is a reasonable value for $T_{\perp 1}$. Thus, if $H^-$ ions start from P1 with perpendicular energy $\approx 2.65$ eV, they will have an energy approximately $\approx 2.5$ eV at the location of the PSMS probe forming a beam current. This is a possible scenario by which the $H^-$ ions could be energized to form a beam. Evidence of the beam characteristics can be verified through analytical fitting of the measured IEDF.

## Appendix B

## Estimation of Positive and Negative Ion Densities from Langmuir Probe and MBMS Measurements

$$\text{For } H^+: Mass - m, Velocity - v_1, Density - n_1, Counts - C_1$$

$$\text{For } H_2^+: Mass - 2m, Velocity - v_2, Density - n_2, Counts - C_2$$

$$\text{For } H_3^+: Mass - 3m, Velocity - v_3, Density - n_3, Counts - C_3$$

$$A = Area\ of\ detection$$

$$T_e = Electron\ Temperature$$

Now, considering the Bohm velocity of positive ions:

$$v_1 = \sqrt{\frac{KT_e}{m}}, v_2 = \sqrt{\frac{KT_e}{2m}}, v_3 = \sqrt{\frac{KT_e}{3m}}$$

Now, for MBMS measurements, the measured intensity (counts/s) can be written as:

$$C_1 = n_1 v_1 A$$

$$C_2 = n_2 v_2 A$$

$$C_3 = n_3 v_3 A$$

Where, $C_1, C_2$ & $C_3$ are the counts corresponding to peak energy of IEDF.

$C_1$ can also be written as:

$$C_1 = n_1 A \sqrt{\frac{T_e}{m}} \quad (1)$$

Which gives:

$$n_1 = \frac{C_1}{A}\sqrt{\frac{m}{T_e}} \quad (2)$$

Similarly,

$$n_2 = \frac{C_2}{A}\sqrt{\frac{2m}{T_e}} \tag{3}$$

$$n_3 = \frac{C_3}{A}\sqrt{\frac{3m}{T_e}} \tag{4}$$

Eq. (2)/(3) gives:

$$\frac{n_1}{n_2} = \frac{C_1}{C_2} \times \frac{1}{\sqrt{2}}$$

$$\Rightarrow n_2 = \frac{C_2}{C_1} \times \sqrt{2} n_1 \tag{5}$$

Similarly,

$$n_3 = \frac{C_3}{C_1} \times \sqrt{3} n_1 \tag{6}$$

Now, from LP measurements, the ion saturation current can be written as:

$$I_{sat} = 0.61\ e\ A_p (n_1 v_1 + n_2 v_2 + n_3 v_3)$$

$$J = \frac{I_{sat}}{0.61 e A_p} = n_1 v_1 + n_2 v_2 + n_3 v_3$$

$$J = n_1 v_1 + n_2 v_2 + n_3 v_3 \tag{7}$$

Substituting the values of $n_2$ *and* $n_3$ from Eq. (5) and (6) in (7) we get:

$$J = n_1 \sqrt{\frac{T_e}{m}} \left(1 + \frac{C_2}{C_1} + \frac{C_3}{C_1}\right)$$

Which gives:

$$n_1 = \frac{J\sqrt{\frac{m}{T_e}}}{\left(1 + \frac{C_2}{C_1} + \frac{C_3}{C_1}\right)} \tag{8}$$

Once $n_1$ is calculated, $n_2$ & $n_3$ can be determined using Eq. (5) and (6).

Now talking about MBMS measurement for $H^-$ ions:

$$C_- = n_- v_- A$$

$$C_- = n_- \sqrt{\frac{2E}{m}}\ A$$

Where, $E$ is the peak energy of $H^-$ IEDF and $C_-$ is the corresponding counts.

$$\Rightarrow n_- = C_-\sqrt{\frac{m}{2E}} \times \frac{1}{A} \tag{9}$$

Dividing Eq. (9)/(2) we get:

$$\frac{n_-}{n_1} = \frac{C_-}{C_1}\sqrt{\frac{T_e}{2E}} \tag{10}$$

Similarly:

$$\frac{n_-}{n_2} = \frac{C_-}{C_2}\sqrt{\frac{T_e}{4E}} \tag{11}$$

$$\frac{n_-}{n_3} = \frac{C_-}{C_3}\sqrt{\frac{T_e}{6E}} \tag{12}$$

Eq. (10) or (11) or (12) can be used to determine the H- density.

## Values taken for the calculation reported in paper:

For Cylindrical Langmuir Probe:

$$I_{sat} = 250 \times 10^{-6} \text{ A}$$
$$Probe\ Diameter, d = 0.25 \text{ mm} = 0.25 \times 10^{-3} \text{ m}$$
$$Probe\ Length, L = 5 \text{ mm} = 5 \times 10^{-3} \text{ m}$$
$$T_e = 2.0 \text{ eV}$$

For PSMS Probe:

$$C_1(H^+) = 712800$$
$$C_2(H_2^+) = 915800$$
$$C_3(H_3^+) = 1865400$$
$$C_-(H^-) = 45800$$
$$E_- = 2.5 \text{ eV}$$